\documentclass[12pt]{article}

\usepackage{graphicx,amsmath}
\usepackage{here}

\begin{document}

\title{\bf Hadronization in Polarized Semi-Inclusive DIS:
the Question of Independent Fragmentation}

\author{Aram~Kotzinian \\[1.cm] \it
 Dipartimento di Fisica Generale, Universit\`a di Torino \\
         and \it INFN, Sezione di Torino, Via P. Giuria 1, I-10125 Torino, Italy\\
and \it Yerevan Physics Institute, 375036 Yerevan, Armenia \\
and \it JINR, 141980 Dubna, Russia\\
\it email: aram.kotzinian@cern.ch}

\maketitle

\begin{abstract}
The new formalism for description of (un)polarized semi-inclusive
DIS at moderate energies is developed. Hadron production is
modeled as a product of distribution functions and hadronization
functions (HFs) weighted by the hard scattering cross sections, as
it is done in the {\tt LEPTO} event generator. The correct
treatment of polarization effects shows that description of
semi-inclusive DIS within this formalism includes a new
nonperturbative input -- {\it polarized} HFs. It is shown that
this approach does not correspond to the commonly adopted one with
the independent quark fragmentation. The purity method used by the
HERMES collaboration mixes up the two approaches and ignores the
contributions from polarized HFs. This method cannot be considered
a precise tool for the extraction of polarized quark distributions
from measured SIDIS asymmetries.
\end{abstract}

\section{Introduction\label{sec:intro}}

Deep inelastic scattering (DIS) is one of the main sources of our
knowledge of the nucleon structure. More information about both
the nucleon structure and a hadron production mechanism one can
obtain by studying the semi-inclusive DIS (SIDIS).

It is evident that the theoretical description of SIDIS is much
more complicated than that of DIS owing to our poor knowledge of
the nonperturbative hadronization mechanism. Traditionally, one
distinguishes two regions for hadron production: the  current
fragmentation region, $x_F>0$ and the target fragmentation region,
$x_F<0$\footnote{We use the standard SIDIS notations and variables
like in~\cite{herm}.}. The  common assumption is that hadrons in
the current fragmentation region with  $z>0.2$  are produced in
the independent quark fragmentation. Then, in the LO approximation
of perturbative QCD the SIDIS cross section for unpolarized target
is given as

\begin{equation}
  \label{eq:sidis}
  \sigma^{h}(x,z,Q^2) \propto (1+(1-y)^2)
  \sum_{q} e_{q}^2 \, q(x,Q^2)\, D_{q}^h(z,Q^2) \,
\end{equation}
and for polarized beam and target
\begin{equation}
  \label{eq:psidis}
  \Delta \sigma^{h}(x,z,Q^2) \propto (1-(1-y)^2)
  \sum_q e_q^2 \, \Delta q(x,Q^2)\, D_q^h(z,Q^2)\, .
\end{equation}
The virtual photon asymmetry for hadron $h$ production can be
expressed as
\begin{equation}
  \label{eq:a1}
  A_1^{h}(x,z,Q^2) =
  \frac{\sum_q e_q^2 \, \Delta q(x,Q^2)\, D_q^h(z,Q^2)}
  {\sum_{q} e_{q}^2 \, q(x,Q^2)\, D_{q}^h(z,Q^2)} \, .
\end{equation}
This equation can be rewritten as follows:
\begin{equation}
  \label{eq:a1pur}
  A_1^{h}(x,z,Q^2)=
  \sum_q {\cal P}_q^h(x,z,Q^2) \, \frac{\Delta q(x,Q^2)}{q(x,Q^2)} \, ,
\end{equation}
where the quark polarizations ($\Delta q / q$) are factored out
and the \textit{purities}, $\mathcal{P}_q^h$, are defined as
\begin{equation}
  \label{eq:pur}
  {\cal P}_q^h(x,z,Q^2) = \frac{e_q^2 \, q(x,Q^2) \, D_q^h(z,Q^2)}
  {\sum_{q'} \, e_{q'}^2 \, q'(x,Q^2) \, D_{q'}^h(z,Q^2)} \, .
\end{equation}

Recently, the important issue of the extraction of polarized quark
distributions was again addressed by the HERMES
collaboration~\cite{herm}. They have used the above LO description
of SIDIS and calculated purities using the Monte Carlo unpolarized
event generator {\tt LEPTO}~\cite{lepto}. Then, using the
asymmetries, measured for different hadrons the helicity
distributions $\Delta q(x)$ were extracted by solving
Eq.~(\ref{eq:a1pur}).

The main assumption of this method is that the hadronization
mechanism in {\tt LEPTO} is the same as in na{\"{\i}}ve picture of
SIDIS where all hadrons in the current fragmentation region with
$z>0.2$ are produced in independent quark fragmentation and there
are no additional terms in both the numerator and the denominator
of Eq.~(\ref{eq:a1}). This assumption is based on factorization
theorem of QCD which is proven as an asymptotic statement for very
high momentum transfers and hence very high energies.

An alternative approach is adopted in {\tt LEPTO}~\cite{lepto}
event generator. The hadronization mechanism of this generator is
based on LUND string fragmentation model implemented in the {\tt
JETSET} program~\cite{JETSET}. In this model the QCD confinement,
the quantum numbers and energy-momentum conservation are taken
into account. As a consequence at moderate beam energies when the
final hadronic system has a limited invariant mass, $\sim 3-5
\;GeV$, one cannot neglect the influence of the target remnant
state on distributions of hadrons produced in the current
fragmentation region.

In Sec.~\ref{sec:lepff} of this paper it will be demonstrated that
the properties of the "quark fragmentation functions" extracted
from generated {\tt LEPTO} samples are in contradiction with
generally accepted properties of independent quark fragmentation.
The reason for this discrepancy is indicated. The generalization
of the parton model expression for polarized SIDIS is given in
Sec.~\ref{sec:polstr}. Finally, in Sec.~\ref{sec:concl} some
discussion and conclusions are presented.

\section{LEPTO and Fragmentation Functions\label{sec:lepff}}

In the standard picture of SIDIS (see Eq.~(\ref{eq:sidis})) the
quark fragmentation functions by definition depend on the type of
hadron, quark flavor and fraction of quark energy carried by
hadron, $z$, (there is also a weak dependence on $Q^2$ due to
perturbative QCD effects), and {\it are independent} of {\bf a)}
the Bjorken variable $x$ and {\bf b)} the target type. These
properties related to universality of fragmentation functions are
essential -- they indicate that one is dealing with independent
quark fragmentation and that there is no influence of the target
remnant on hadron production in the current fragmentation region.

These fragmentation functions are not well known for different
hadron and quark types and the {\tt LEPTO} event generator is used
by HERMES collaboration~\cite{herm} to calculate purities.
However, hadronization in this generator is based on the LUND
string fragmentation model and one has first to check if the
issues {\bf a)} and {\bf b)} are satisfied in this approach. To
this end samples of SIDIS events were generated for HERMES
experimental conditions using the settings of {\tt LEPTO} as
in~\cite{herm} (see also reference [64] of~\cite{herm}). The
option {\tt LST(8)=0} of {\tt LEPTO} was used since it corresponds
to the LO approximation of SIDIS \footnote{The option {\tt
LST(8)=1} used in reference [64] of~\cite{herm} corresponds to
inclusion in the generation of the first order QCD matrix elements
for gluon radiation and photon-gluon fusion. With this option
Eq.~(\ref{eq:sidis}) has to be replaced with the NLO expression to
include the gluon distribution and fragmentation functions which
are not involved in the LO purity analysis. However, it was
checked that the conclusions of the present paper remain valid
independently of the choices of the {\tt LEPTO} settings at HERMES
energy.}. The following cuts are used: $Q^2>1\, GeV^2$, $W^2>10
\,GeV^2$, $y<0.85$, $0.023<x<0.6$, $E'>3.5 \,GeV$, $z>0.2$ and
$x_F>0.1$.

\begin{figure}[h!]
\begin{center}
\vspace {-0.5cm}
 \includegraphics[width=0.95\linewidth]{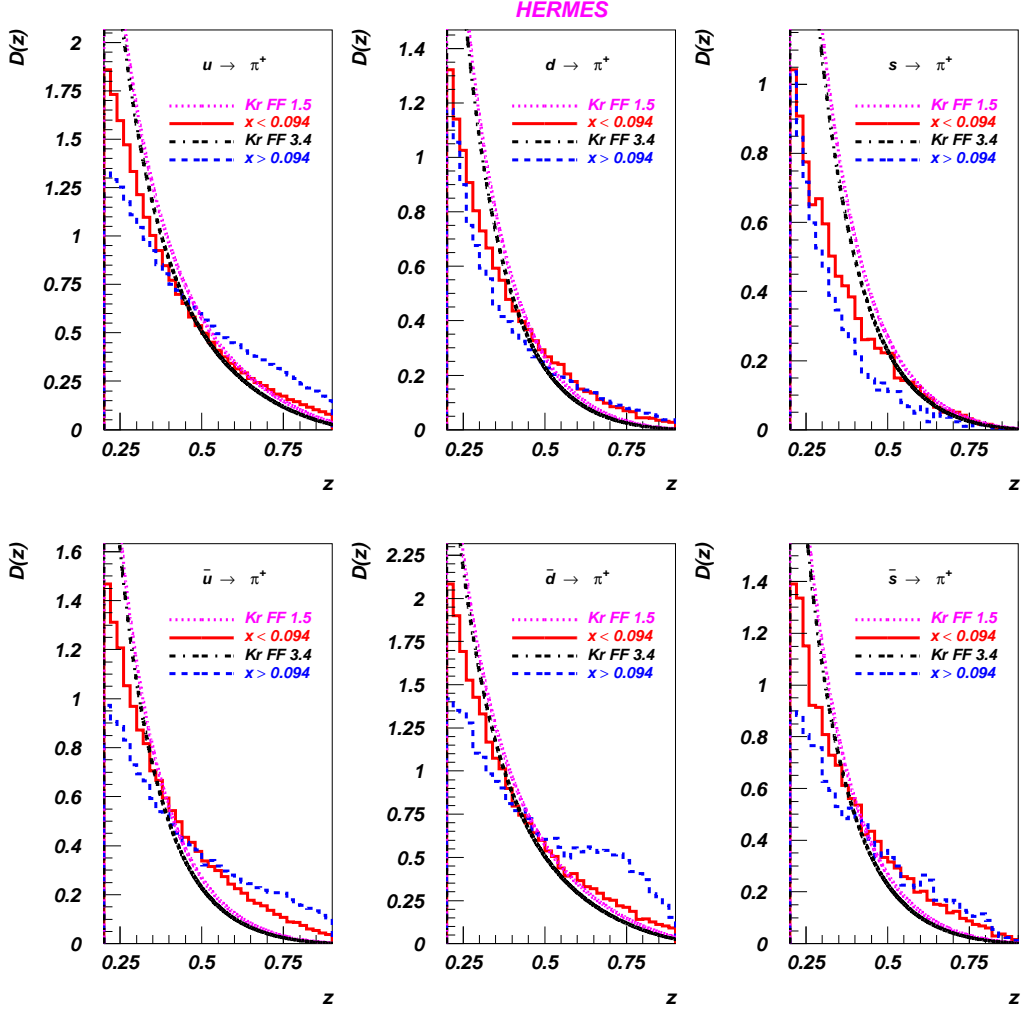}
\vspace {-0.7cm} \caption{\label{fig:pipherm} {\it "Quark
fragmentation functions" to $\pi^+$ calculated for HERMES
conditions. Solid line -- with cut $x<0.094$, dashed line -- with
cut $x>0.094$. Fragmentation functions from Kretzer: dotted line
-- for $Q^2=1.5 (GeV/c)^2$, dot-dashed line -- for $Q^2=3.4
(GeV/c)^2$.}}
\end{center}
\vspace {-0.5cm}
\end{figure}

In Fig.~\ref{fig:pipherm} the quark "fragmentation functions" to
$\pi^+$, extracted from the sample, generated for the HERMES
kinematics on a proton target, are presented as a function of $z$.
The available range of the Bjorken variable, $x$, was divided into
two equally populated intervals: {\bf 1)} ($x<0.094$) and {\bf 2)}
($x>0.094$). As one can see from Fig.~\ref{fig:pipherm}, the
extracted "quark fragmentation functions" happen to be strongly
dependent on the Bjorken $x$ variable in striking contradiction
with the property {\bf a)} mentioned above. Note, that this cannot
be attributed to the (weak) $Q^2$-dependence of fragmentation
functions. To demonstrate this the LO fragmentation functions
from~\cite{kretzer}, which includes the QCD evolution, are also
presented in the same figure for mean values of $Q^2$
corresponding to Bjorken variable intervals {\bf 1)} ($Q^2=1.5
(GeV/c)^2$) and {\bf 2)} ($Q^2=3.4 (GeV/c)^2$).

\begin{figure}[h!]
\begin{center}
\vspace {-0.5cm}
 \includegraphics[width=0.95\linewidth]{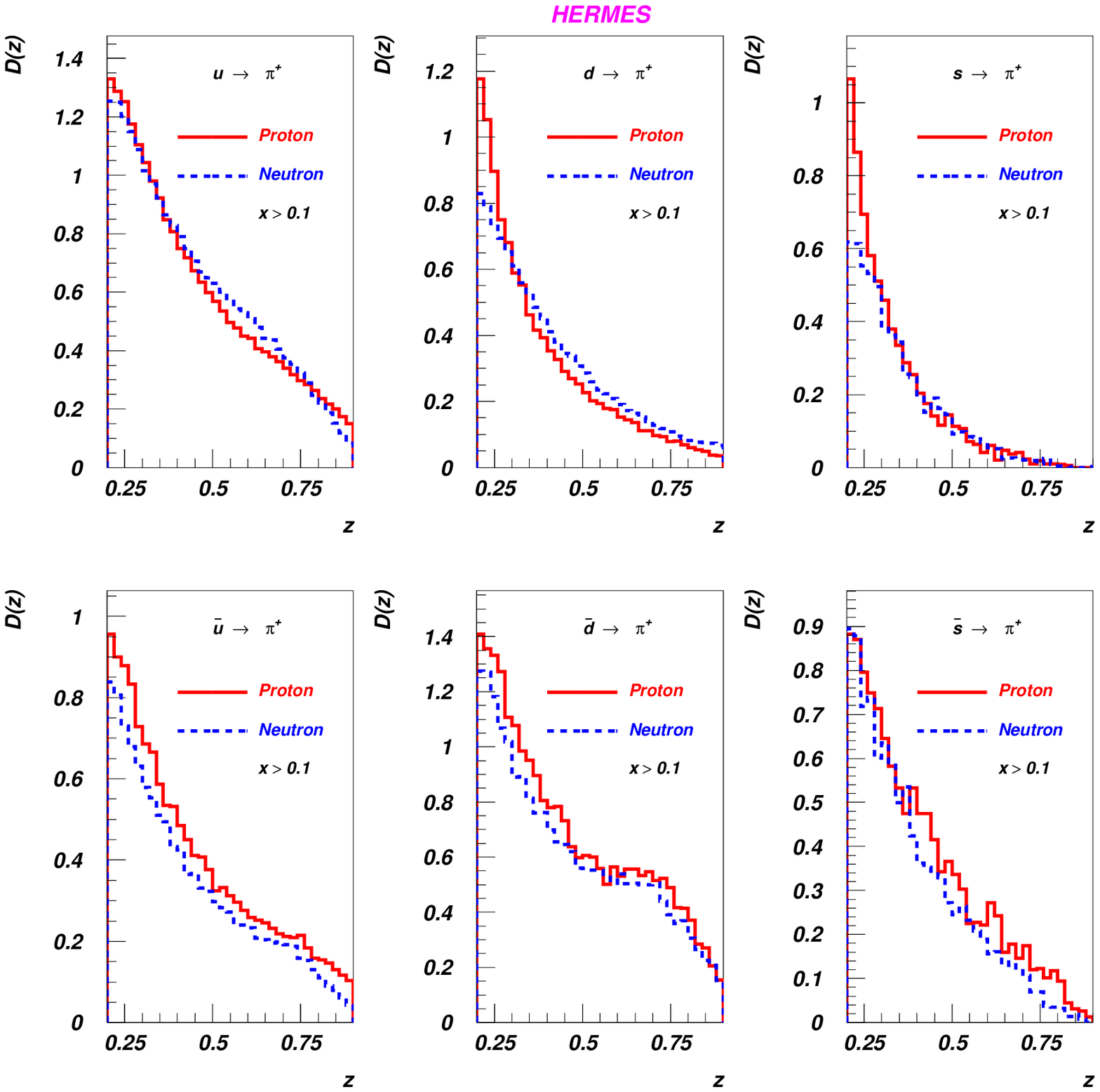}
\vspace {-0.7cm} \caption{\label{fig:prntrherm} {\it "Quark
fragmentation functions" to $\pi^+$ calculated for HERMES
conditions with cut $x>0.1$. Solid line -- proton target, dashed line
-- neutron target.}}
\end{center}
\vspace {-0.5cm}
\end{figure}

The "quark fragmentation functions" obtained from the samples
generated for proton and neutron targets with the cut $x>0.1$ are
presented in Fig.~\ref{fig:prntrherm}. We see a dependence on the
target type which contradicts property {\bf b)} of fragmentation
functions.

Such behavior of "fragmentation functions" extracted from the
generated samples is also observed for the production of other
types of light meson like $\pi^{-}$,  $K^{+}$, $K^{-}$ {\it etc}.

At this point one can conclude that the "quark fragmentation
functions" extracted from the samples generated for HERMES
kinematical conditions do not correspond to the commonly used
notion of fragmentation function.

Let us recall that hadronization in the {\tt LEPTO} event
generator is based on string fragmentation and as it is stressed
in~\cite{JETSET}: {\it "the primary hadrons produced in string
fragmentation come from the string as a whole, rather than from an
individual parton"}. In other words the distributions of the
produced hadrons "retains the memory" not only of the struck quark
type but also of the target remnant and, hence, the entire string
configuration.

Event generation in {\tt LEPTO} proceeds in three steps: first the
hard scattering kinematics ($x,Q^2$) is chosen from the differential
DIS cross-section. Second, the struck quark flavor is chosen.
Third, the string is set up and hadronized according to LUND model
implemented in {\tt JETSET} program~\cite{JETSET}. Within this
approach the SIDIS cross section can be expressed as

\begin{equation}
  \label{eq:sidisstr}
  \sigma^{h}_{N}(x,z,Q^2) \propto (1+(1-y)^2)
  \sum_{q} e_{q}^2 \, q(x,Q^2)\, H_{q/N}^h(x,z,Q^2) \, ,
\end{equation}
where the functions $H_{q/N}^h(x,x_F,Q^2)$ are describing the {\it
conditional} probability of hadron $h$ production in the
hadronization of the system formed by struck quark $q$ and
corresponding target remnant. Let us call them {\it hadronization
functions}, HFs. The target remnant type and, hence, the whole
fragmenting system configuration depends both on the nucleon type
and on the struck quark type\footnote{The fragmenting system in
{\tt LEPTO/JETSET} is in general more complicated than a
quark-diquark string. The target remnant state depends on the
removed active parton type and the whole fragmenting system may
contain multi-string configurations~\cite{lepto,JETSET}.}. This
means that in contrast with independent fragmentation functions
the HFs are non universal -- they depend on the process type and
energy.

Note, that Eq.~(\ref{eq:sidisstr}) is valid not only in the
current fragmentation region but in the whole $x_F$ interval.

The product $q(x,Q^2)\, H_{q/N}^h(x,z,Q^2)$ is the probability to
find the parton $q$ in the nucleon, $N$, and, after hard
interaction, to create a hadron $h$ in the string hadronization.
By its physical meaning (probabilistic interpretation) this object
represents nothing else but the fracture functions discussed
in~\cite{tv}. There exist certain arguments based on handbag
diagram dominance that this concept may be applied even in the
current fragmentation region of SIDIS~\cite{teryaev}. The {\tt
LEPTO/JETSET} Monte Carlo program can be considered as a model for
these functions. It is clear that from the generated samples one
can actually  extract only HFs and as one can see from
Fig.~\ref{fig:pipherm} and Fig.~\ref{fig:prntrherm} even in the
current fragmentation region one cannot neglect the dependence of
this functions on the Bjorken x variable and on the target type.

\begin{figure}[h!]
\begin{center}
\vspace {-0.5cm}
 \includegraphics[width=0.95\linewidth]{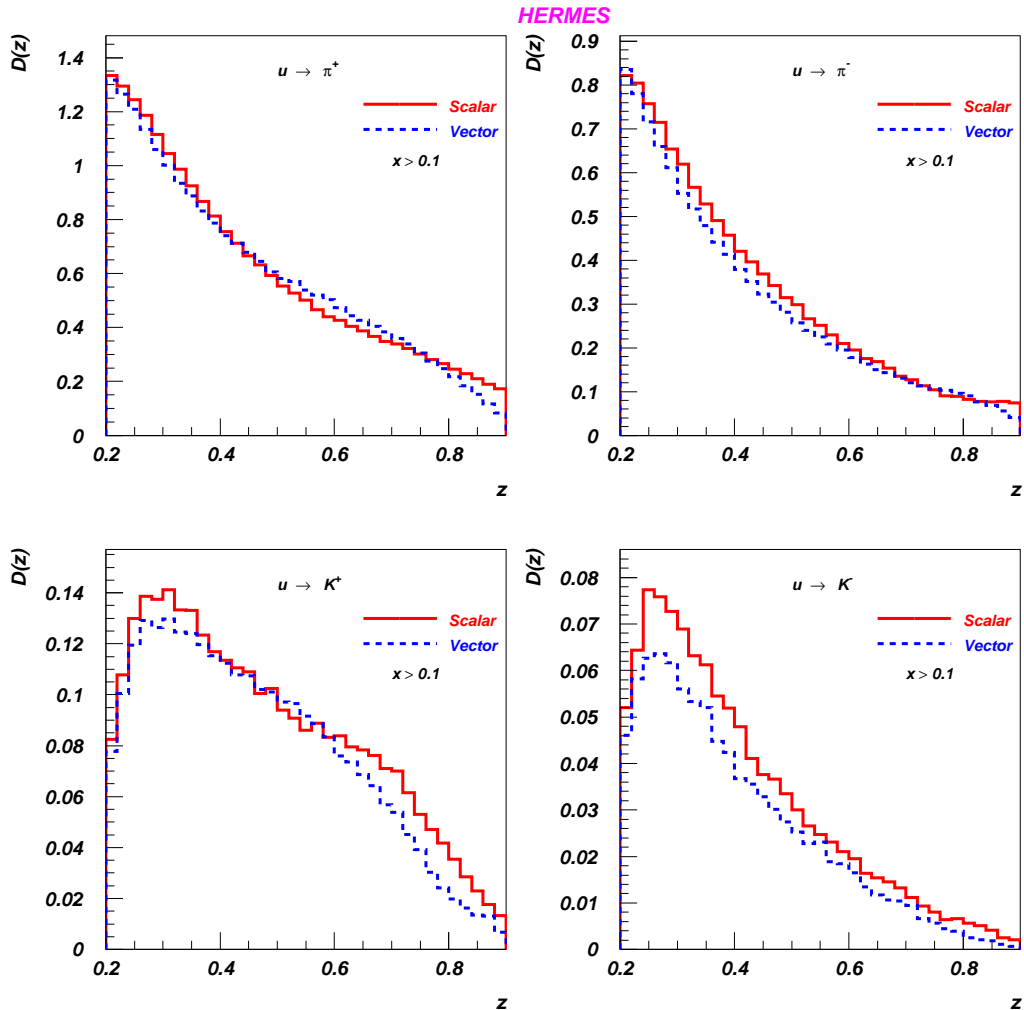}
\vspace {-0.7cm} \caption{\label{fig:diqrkherm} {\it "Quark
fragmentation functions" dependence on diquark type calculated for
HERMES conditions with cut $x>0.1$. Solid line -- scalar diquark,
dashed line -- vector diquark.}}
\end{center}
\vspace {-0.5cm}
\end{figure}

As a further consideration of the target remnant state influence
on the "quark fragmentation functions" let us now consider the
spin inside LEPTO. In the simplest case when the valence $u$-quark
is removed by hard scattering from a proton, the target remnant is
a scalar, $(ud)_0$, diquark with probability $w_0=0.75$ or a
vector, $(ud)_1$, diquark with relative probability
$w_1=1-w_0=0.25$~\cite{lepto, JETSET}. In Fig.~\ref{fig:diqrkherm}
the "fragmentation functions" for $h=\pi^{+}, \pi^{-}, K^{+}$ and
$K^{-}$ extracted with $x>0.1$ cut are presented for the cases
when the target remnant diquark is chosen to be a 100\% scalar
($w_0=1$) or a vector one ($w_0=0$). We see that HFs calculated
with our generated samples exhibit dependence (at 5--10\% level)
on the target remnant spin state already in unpolarized SIDIS.

This dependence has a quite general origin. To understand the
underlying physics let us consider, as an example, a $K^{+}$ meson
production. At some stage of hadronization the $\bar s s$-pair is
created and $s$-quark can combine with the remnant diquark to form
a strange baryon. The formation of first rank $\Lambda$ hyperon is
possible for a scalar diquark remnant and forbidden for the case
of vector diquark. In the second case only a heavier strange
hyperon can be formed at first rank. Then, due to the
energy--momentum conservation the available energy for a $K^{+}$
meson production will be less in the second case compared with the
first case. This is the reason why in Fig.~\ref{fig:diqrkherm} the
"$u$-quark fragmentation functions into $K^{+}$" is higher for
scalar diquark sample.

\section{Polarized SIDIS and String Fagmentation\label{sec:polstr}}

In the ordinary factorized picture the same {\it unpolarized}
fragmentation function is entering into expression for the cross
section of unpolarized and polarized SIDIS. Let us now compare
that what will happen if we include spin and generalize the
picture with the HFs for SIDIS.

At present there is no the polarized version of the string
fragmentation Monte-Carlo program for event generation. It is
evident that the description of polarized SIDIS is more
complicated than in the unpolarized case. Let us, as an example,
again consider the simplest case when the valence $u$-quark with
positive ($u^{+}$) or negative ($u^{-}$) helicity is removed from
nucleon with positive or negative helicity, $N^{+}$ or $N^{-}$.
Within the SU(6) quark-diquark model used in~\cite{lepto, JETSET}
the polarized nucleon wave functions are given as

\begin{equation}
  \label{eq:polpr}
  p^{+}=\frac{1}{\sqrt{18}}\{u^{+}[3(ud)_{0,0}+(ud)_{1,0}]-\sqrt{2}u^{-}(ud)_{1,1}
  -\sqrt{2}d^{+}(uu)_{1,0}+2d^{-}(uu)_{1,1}\},
\end{equation}
\begin{equation}
  \label{eq:polntr}
  n^{+}=\frac{1}{\sqrt{18}}\{d^{+}[3(ud)_{0,0}+(ud)_{1,0}]-\sqrt{2}d^{-}(ud)_{1,1}
  -\sqrt{2}u^{+}(dd)_{1,0}+2u^{-}(dd)_{1,1}\},
\end{equation}
where $(q_1q_2)_{(i,k)}$ stands for the diquark formed by the
$q_1$- and $q_2$-quarks with spin {\it i} and spin projection {\it
k}.

Using the explicit form of the polarized nucleon wave functions
one can calculate the relative probabilities, $w$, of different
target remnant states (and hence the states of entire string)
depending on the struck quark and the nucleon polarizations. For
example, when the $u^{+}$-quark is removed from the $p^{+}$ we get
the following string configurations with corresponding
probabilities $w$
\begin{equation}
  \label{eq:strp+}
  p^{+} \ominus u^{+} \Longrightarrow
  \begin{cases}
  \{(ud)_{0,0}\cdot\cdot\cdot\cdot\cdot u^{+}\}, \qquad  w=0.9, \\
  \{(ud)_{1,0}\cdot\cdot\cdot\cdot\cdot u^{+}\}, \qquad  w=0.1,
  \end{cases}
\end{equation}
where $\{(q_1q_2)_{i,k}\cdot\cdot\cdot\cdot\cdot q^{+}\}$ denotes
the string formed by the struck quark $q^{+}$ and the diquark
$(q_1q_2)_{i,j}$. Similarly, when the $u^{+}$-quark is removed
from $p^{-}$ we get
\begin{equation}
  \label{eq:strp-}
  p^{-} \ominus u^{+} \Longrightarrow
  \{(ud)_{1,-1}\cdot\cdot\cdot\cdot\cdot u^{+}\}, \qquad w=1.
\end{equation}
For the neutron target we have
\begin{equation}
  \label{eq:stntr+1}
  n^{+} \ominus u^{+} \Longrightarrow
  \{(dd)_{1,0}\cdot\cdot\cdot\cdot\cdot u^{+}\}, \qquad w=1,
\end{equation}
and
\begin{equation}
  \label{eq:stntr-1}
  n^{-} \ominus u^{+} \Longrightarrow
  \{(dd)_{1,-1}\cdot\cdot\cdot\cdot\cdot u^{+}\}, \qquad w=1.
\end{equation}

The relations~(\ref{eq:strp+}-\ref{eq:stntr-1}) demonstrate that
the string configuration indeed depends not only on the struck
quark type and its polarization but also on the target type and
polarization. As we have seen in Sec.~\ref{sec:lepff} the
description of hadron production in the current fragmentation
region of SIDIS within the LUND fragmentation model does not
correspond exactly to the commonly adopted simple picture of
independent quark fragmentation but rather to the more complicated
approach based on fracture functions. Even in unpolarized SIDIS
HFs depend on target fragment spin states as it is demonstrated in
Fig.~\ref{fig:diqrkherm}. Then in polarized SIDIS the dependence
on the target and struck quark polarizations appears. So, one has
to generalize Eq.~(\ref{eq:sidisstr}) for the polarized SIDIS
case.

Let us start with the SIDIS cross section $\sigma^{h}_{N \lambda_l
\lambda_N}$ for the positive helicity lepton, $\lambda_l=+1$ and
hadron, $\lambda_N=+1$:

\begin{equation}
  \label{eq:psidis++}
  \sigma^{h}_{N++} \propto
  \sum_{q} e_{q}^2 \{ q^{+}\, H_{q/N++}^h + (1-y)^2q^{-}\, H_{q/N-+}^h\},
\end{equation}
where $H_{q/N\lambda_q \lambda_N}^h$ describes the production
probability of the hadron $h$ in the quark--target remnant system
fragmentation and depends not only on $x$ and $z$ (or $x$ and
$x_F$) but also on the struck quark and nucleon helicities,
$\lambda_q=\pm 1$ and $\lambda_N=\pm 1$. Similarly
\begin{equation}
  \label{eq:psidis+-}
  \sigma^{h}_{N+-} \propto
  \sum_{q} e_{q}^2 \{ q^{-}H_{q/N+-}^h + (1-y)^2q^{+}H_{q/N--}^h\}.
\end{equation}

The partially polarized beam state, $l^{\lambda_l}$, (with
helicity $\lambda_l$) can be described as
$l^{\lambda_l}=1/2(1+\lambda_l)l^{+}+1/2(1-\lambda_l)l^{-}$ and
similarly for the nucleon. Then in the general case of arbitrary
polarized beam and target we have
\begin{eqnarray}
  \label{eq:arbpol}
  \sigma^{h}_{N \lambda_l \lambda_N} \propto
  (1+\lambda_l)(1+\lambda_N)\sigma^{h}_{N++}+
  (1+\lambda_l)(1-\lambda_N)\sigma^{h}_{N+-}+& \nonumber \\
  (1-\lambda_l)(1+\lambda_N)\sigma^{h}_{N-+}+
  (1-\lambda_l)(1-\lambda_N)\sigma^{h}_{N--}. \;\; &
\end{eqnarray}
Now, using the relations $H_{q/N++}^h = H_{q/N--}^h$ and
$H_{q/N+-}^h = H_{q/N-+}^h$ which follow from parity invariance
and introducing
\begin{eqnarray}
 H_{q/N}^h = H_{q/N++}^h + H_{q/N+-}^h \nonumber \\
 \Delta H_{q/N}^h = H_{q/N++}^h - H_{q/N+-}^h
\end{eqnarray}
after simple algebra one get:

\begin{eqnarray}
  \label{eq:psidisfrac}
  \sigma^{h}_{N \lambda_l \lambda_N} \propto [1+(1-y)^2]
  \sum_{q} e_{q}^2 \{ q H_{q/N}^h + \Delta q \Delta H_{q/N}^h\}+&\nonumber \\
  \lambda_l \lambda_N [1-(1-y)^2]
  \sum_{q} e_{q}^2 \{ \Delta q H_{q/N}^h + q \Delta H_{q/N}^h\}, \;\; &
\end{eqnarray}
where now $\lambda_l$ and $\lambda_N$ are the (arbitrary) beam and
target helicities.

Eq.~(\ref{eq:psidisfrac}) is very similar to the equation proposed
in~\cite{gr}. The difference is that functions $H_{q/N}^h$ and
$\Delta H_{q/N}^h$ are not independent quark fragmentation
functions like in~\cite{gr}. It is well known that the single spin
dependence is forbidden by the parity invariance for integrated
over transverse momentum independent fragmentation. By this reason
the same quark fragmentation functions enter in
Eqs.~(\ref{eq:sidis}-\ref{eq:psidis}).In contrast, HFs describe
the probability of hadron production in the hadronization of the
whole {\it struck quark -- target remnant} system and, hence, one
deal with the double spin effects. Even for hadrons produced in
the current fragmentation region these functions can depend not
only on the fraction of the quark energy carried by produced
hadron but also on the whole hadronic CMS energy and {\it the
target and the struck quark polarizations}.

When integrated over the whole available phase space of the
selected hadron, Eq.~(\ref{eq:psidisfrac}) transforms to the
standard parton model expression for the polarized DIS, provided
that the following sum rules hold valid:

\begin{eqnarray}
  \label{eq:sumrul}
    &\sum_h\int dz\,z H_{q/N}^h(x,z,Q^2) = 1 ,\nonumber \\
    &\sum_h\int dz\,z \Delta H_{q/N}^h(x,z,Q^2)=0.
\end{eqnarray}

\section{Discussion and Conclusions\label{sec:concl}}

The standard expression for the SIDIS description in the current
fragmentation region\footnote{Note that these expressions are
based on QCD factorization theorems, which represent asymptotic
statements valid for very high  lepton energies, $Q^2$ and $W$.
Only under these conditions can one neglect the influence of the
target remnant on hadron production in the current fragmentation
region.} is obtained if one assume that
\begin{equation}
  \label{eq:ff}
  H_{q/N}^h(x,z,Q^2) \rightarrow D_q^h(z,Q^2)
\end{equation}
 and
\begin{equation}
  \label{eq:pff}
  \Delta H_{q/N}^h(x,z,Q^2) \rightarrow 0.
\end{equation}

\begin{figure}[h!]
\begin{center}
\vspace {-0.5cm}
 \includegraphics[width=0.95\linewidth]{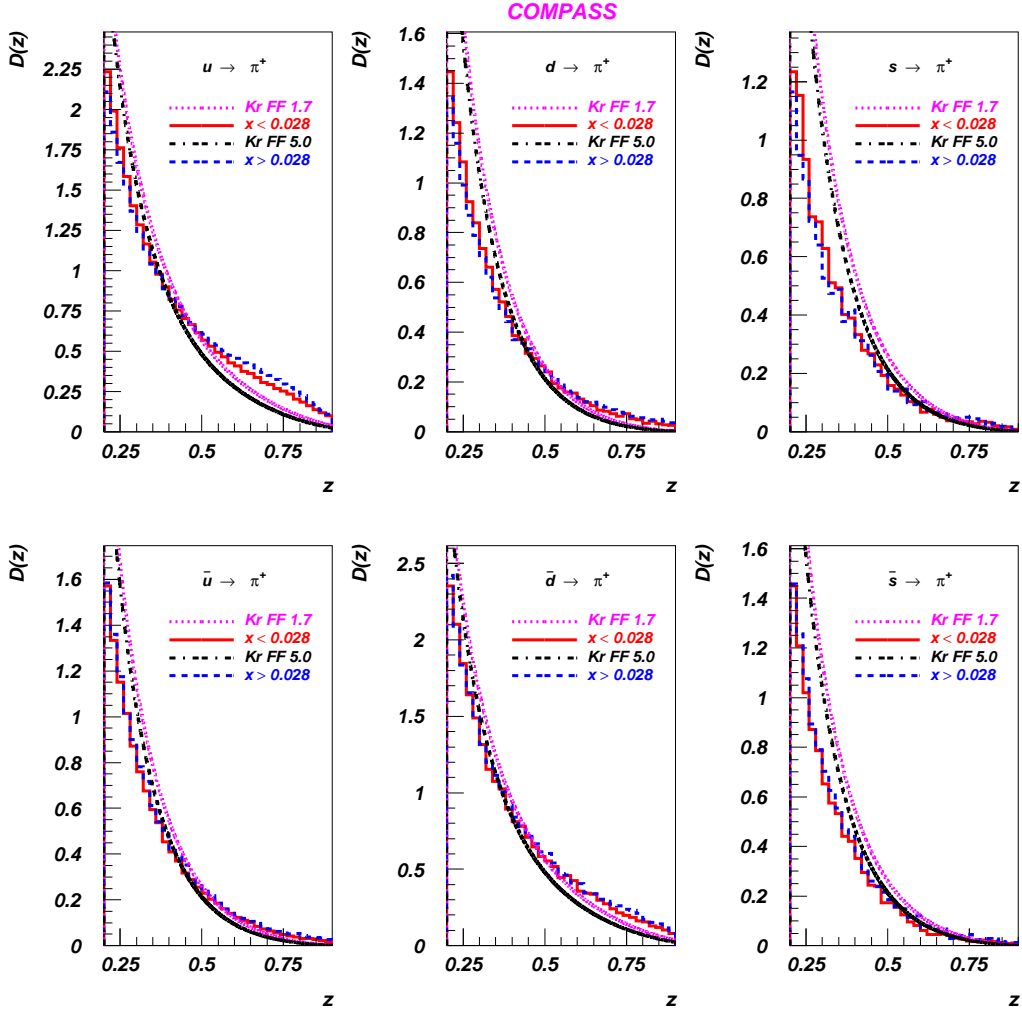}
\vspace {-0.7cm} \caption{\label{fig:pipcomp} {\it "Quark
fragmentation functions" to $\pi^+$ calculated for COMPASS
conditions. Solid line -- with the $x<0.028$, dashed line -- with
cut $x>0.028$. Fragmentation functions from Kretzer: dotted line
-- for $Q^2=1.7 (GeV/c)^2$, dot-dashed line -- for $Q^2=5.0
(GeV/c)^2$.}}
\end{center}
\vspace {-0.5cm}
\end{figure}

As we have demonstrated in Sec.~\ref{sec:lepff}
relation~(\ref{eq:ff}) is not correct for the HERMES experimental
conditions in the LUND fragmentation model. On the other hand we
have seen in Fig.~\ref{fig:diqrkherm} that hadronization in this
model depends on the target remnant spin quantum numbers. In the
case of polarized SIDIS the relative probabilities of different
target remnant states depend on the target and struck quark
polarizations, see Eqs.~(\ref{eq:strp+}-\ref{eq:strp-}). So, there
is no any reason to believe that relation~(\ref{eq:pff}) will hold
for the polarized SIDIS at moderate energies. Thus, a {\it new
nonperturbative inputs -- the polarized HFs}, $\Delta
H_{q/N}^h(x,z,Q^2)$, are needed. In this case Eq.~(\ref{eq:a1})
underlying the purity method is not exact and must to be replaced
by
\begin{equation}
  \label{eq:a1fracf}
  A_1^{h}(x,z,Q^2) =
  \frac{\sum_q e_q^2 \,q(x,Q^2)H_{q/N}^h(x,z,Q^2)
  [\frac{\Delta q(x,Q^2)}{q(x,Q^2)}+
  \frac{\Delta H_{q/N}^h(x,z,Q^2)}{H_{q/N}^h(x,z,Q^2)}]}
  {\sum_{q} e_{q}^2 \,q(x,Q^2)\,H_{q/N}^h(x,z,Q^2)
  [1+\frac{\Delta q(x,Q^2)\,
  \Delta H_{q/N}^h(x,z,Q^2)}{q(x,Q^2)\,H_{q/N}^h(x,z,Q^2)}]} \, ,
\end{equation}
with extra contributions in the numerator and denominator as
compared to Eq.~(\ref{eq:a1}). As one can see from
Fig.~\ref{fig:diqrkherm} the second term in the square brackets in
numerator can reach 5--10\%. Its influence can still be negligible
in the unpolarized cross section (second term in square brackets
of denominator), since it entering multiplied to quark
polarization. Whereas in numerator it can be comparable with (or
even greater than) the first term.

Though we are not able to calculate from the first principles the
HFs, one can try to estimate the possible effects of polarized HFs
on the extraction of polarized quark distribution. For example,
one could estimate the effects of scalar and vector diquarks using
formalism of Sec.~\ref{sec:lepff} and calculate $\frac{\Delta
H_{q/N}^h(x,z,Q^2)}{H_{q/N}^h(x,z,Q^2)}$ for each $x$-bin.

In~\cite{ak} a model for the extra contribution in the numerator
of Eq.~(\ref{eq:a1fracf}) has been considered and it was
demonstrated that our ignorance of polarized HFs may lead to
incorrect results for polarized sea quark distributions. Here it
is demonstrated that using {\tt LEPTO} in analysis based on the
factorized approach is inconsistent. As a consequence it is
essential in using factorization approach to polarized quark
distribution extraction first to check it, since the polarization
dependence is more sensitive to factorization that the unpolarized
multiplicity distributions.

It is interesting to study how these effects depend on the energy
and in particular how they will affect the COMPASS~\cite{comp}
analysis. In Fig.~\ref{fig:pipcomp} the same distributions as in
Fig.~\ref{fig:pipherm} are presented for COMPASS kinematics. One
can see that the dependence on Bjorken variable of the "quark
fragmentation functions" extracted from generated samples is less
pronounced that in case of HERMES experimental conditions. The
same observation is also valid for the dependence upon the target
type and the target remnant spin state dependencies. This means
that that polarized HFs may be negligible in the current
fragmentation region at high energies.

The string configurations considered
in~(\ref{eq:strp+}-\ref{eq:stntr-1}) correspond to the simplest
case of removing the valence quark from nucleon. In the case, when
the virtual photon interacts with the sea quark or higher order
hard scattering processes are considered, the target remnant and
the final parton configuration are more complicated~\cite{lepto}.
For example, in photon-gluon fusion the target remnant is split
into a quark and a diquark that form two respective separate
strings with the antiquark and quark produced in the fusion
process. One can generalize Eq.~(\ref{eq:psidisfrac}) to include
the corrections from higher order QCD hard processes. The
generalized expression for the polarized cross section of single
and two hadron productions will, in addition, contain new unknown
HFs, $\Delta H_{g/N}^h(x,z,p_T^h,Q^2)$ and $\Delta
H_{g/N}^{h_1,h_2}(x,z_1,z_2,p_T^{h_1},p_T^{h_2},Q^2)$, with the
corresponding distribution functions, $(\Delta) g(x,Q^2)$. This
means that the validity of the Monte Carlo based
approach~\cite{bhk} to extract the polarized gluon distribution
may also be questionable at moderate energies.

It has been recently noted~\cite{teryaev} that the appearance of
separate distribution and fragmentation functions cannot be proven
in general, but is rather assumed and justified {\it a
posteriori}, while the natural framework to describe SIDIS
involves fracture functions. These functions can be also
generalized to describe the T-odd single spin
asymmetries~\cite{teryaev}. As it was mentioned in
Sec.~\ref{sec:lepff} within the LUND fragmentation framework the
fracture functions can be represented by products of distribution
functions and HFs. Recent developments in the theory of SIDIS for
single spin asymmetries and diffractive phenomena also show that
one cannot neglect the interaction of (colored) removed partons
and target remnants (see, for example, ~\cite{brodsky} and
references therein, and ~\cite{teryaev}). This again indicates
that the description of SIDIS based on the na{\"{\i}}ve parton
model with the independent fragmentation is only an approximation,
to be justified at moderate energies.

Let us stress that the approach developed in Sec.~\ref{sec:polstr}
is not directly derived from QCD but is similar to one used in
{\tt LEPTO} event generator. It assumes that in the DIS regime the
hard scattering and the hadronization are factorized. As it was
mentioned long time ago~\cite{berger} the concept of independent
fragmentation can be justified only when there is enough phase
space for the final hadronic system. The important issue is not
only high $Q^2$ but also large rapidity interval available for a
given hadron production. In contrast to the independent
fragmentation model the LUND model deals with the whole final
quark -- target remnant hadronization and takes into account the
energy--momentum conservation, the color flow and the quantum
number correlations. As a consequence, at moderate $W$ there is a
non-negligible influence of the target remnant state on the
distributions of hadrons produced in the current fragmentation
region, as it explained in the end of Sec.~\ref{sec:lepff}.

\section*{Acknowledgements}

The author express his gratitude to M.~Anselmino, A.~Efremov,
E.~Leader and O.~Teryaev for discussions, G.~Pontecorvo for
careful reading of the manuscript and also to NUCLEOFIT group
members of the General Physics Department "A.~Avogadro" of the
Turin University for interest in this work.


\begin{thebibliography}{99}

\bibitem{herm}
  A.~Airapetian {\it et al.}  [HERMES Collaboration],
  Phys.\ Rev.\ D {\bf 71} (2005) 012003, arXiv:hep-ex/0407032.

\bibitem{lepto}
  G.~Ingelman, A.~Edin and J.~Rathsman,
  Comp. Phys. Commun. {\bfseries 101} 108 (1997).

\bibitem{JETSET}
  T.~Sj\"ostrand, Comp. Phys. Commun. {\bfseries 39} 347 (1986),
{\bfseries 43} 367 (1987);\\
T.~Sj\"ostrand, {\it PYTHIA 5.7 and JETSET 7.4: Physics
  and Manual}, arXiv:hep-ph/9508391; \\
 T.~Sj\"ostrand, {\it PYTHIA 6.2: Physics and Manual},
 arXiv:hep-ph/0108264.

\bibitem{kretzer}
S.~Kretzer,
Phys.\ Rev.\ D {\bf 62} (2000) 054001.

\bibitem{tv} L.~Trentadue and G.~Veneziano, Phys. Lett. {\bfseries
  B323} 201 (1994).

\bibitem{teryaev}
O.~V.~Teryaev,
Acta Phys.\ Polon.\ B {\bf 33}, 3749 (2002), arXiv:hep-ph/0211027;\\
Proceedings of X Advanced Research Workshop on High Energy Spin
Physics, Dubna, September 16--20, 2003, p. 200.

\bibitem{gr} M.~Gluck and E.~Reya, arXiv:hep-ph/0203063.

\bibitem{comp}
  The COMPASS Proposal, CERN/SPSLC 96-14, SPSC/P297, March 1996.

\bibitem{ak}
A.~Kotzinian,
Phys.\ Lett.\ B {\bf 552} (2003) 172, arXiv:hep-ph/0211162.

\bibitem{bhk}
A.~Bravar, D.~von Harrach and A.~Kotzinian,
Phys.\ Lett.\ B {\bf 421} (1998) 349, arXiv:hep-ph/9710266.

\bibitem{brodsky}
  S.~J.~Brodsky, R.~Enberg, P.~Hoyer and G.~Ingelman,
  Phys.\ Rev.\ D {\bf 71} (2005) 074020, arXiv:hep-ph/0409119.

\bibitem{berger}
E.~Berger,
Proceedings of the NPAS Workshop on Electronuclear Physics with
Internal Targets (SIAC, 1987), SLAC report 316, eds R G Arnold and
R C Mmehart, p 82; Preprint ANL-HEP-CP-87-45, April 30, 1987.

\end{thebibliography}
\end{document}